\documentclass[a4paper]{jpconf}

\usepackage{graphicx}
\usepackage{amsmath}
\usepackage{bm}
\usepackage{aas_macros}

\begin{document}

\title{The Kelvin-Helmholtz instability and smoothed particle hydrodynamics}

\author{Terrence S Tricco}

\address{Canadian Institute for Theoretical Astrophysics, University of Toronto, 60 St. George St., Toronto, ON, M5S 3H8, Canada}

\ead{ttricco@cita.utoronto.ca}

\begin{abstract}
There has been interest in recent years to assess the ability of astrophysical hydrodynamics codes to correctly model the Kelvin-Helmholtz instability. Smoothed particle hydrodynamics (SPH), in particular, has received significant attention, though there has yet to be a clear demonstration that SPH yields converged solutions that are in agreement with other methods. We have performed SPH simulations of the Kelvin-Helmholtz instability using the test problem put forward by Lecoanet et al (2016). We demonstrate that the SPH solutions converge to the reference solution in both the linear and non-linear regimes. Quantitative convergence in the strongly non-linear regime is achieved by using a physical Navier-Stokes viscosity and thermal conductivity. We conclude that standard SPH with an artificial viscosity can correctly capture the Kelvin-Helmholtz instability.
\end{abstract}

\section{Introduction}

The goal of numerical simulation is to recreate real physics. Some numerical methods may yield a more accurate solution, but the true test of any numerical method is whether it produces solutions that are convergent upon a `true' physical answer with increased computational resources. 

This work studies the convergence properties of SPH (smoothed particle hydrodynamics; \cite{lucy77, gm77}) on the Kelvin-Helmholtz instability. The Kelvin-Helmholtz instability occurs when two fluids move past each other, causing exponential growth of velocity orthogonal to the flow. This develops into vortices which curl back upon themselves -- the hallmark of the instability. As the instability becomes non-linear, secondary instabilities form leading to the two fluids becoming well-mixed along the interface. The Kelvin-Helmholtz instability is inherently a mixing instability.

The ability of SPH, a Lagrangian particle-based method for solving the equations of hydrodynamics \cite{lucy77, gm77}, to correctly model the Kelvin-Helmholtz instability has been questioned in recent years \cite{springel10, hs10, bs12, sijackietal12, hopkins13, hopkins15}. Agertz et al \cite{agertzetal07} concluded that there exists ``fundamental differences between SPH and grid methods'' in their ability to resolve mixing instabilities on the basis of SPH calculations that showed no growth of the Kelvin-Helmholtz instability.

The abysmal results by Agertz et al \cite{agertzetal07} were a consequence of their initial conditions which contained discontinuities in density, velocity and internal energy, combined with a lack of numerical treat for contact discontinuities. Discontinuities require special care numerically. For SPH, velocity discontinuities are treated by an artificial viscosity, and density discontinuities are handled naturally by the density summation since it makes no assumption about differentiability \cite{price08}. No treatment for the initial contact discontinuity was used by Agertz et al \cite{agertzetal07}. It has been demonstrated that including an artificial thermal conductivity is effective in capturing the Kelvin-Helmholtz instability and promoting mixing \cite{price08, wvc08}.

Discontinuous initial conditions preclude formal convergence studies of the Kelvin-Helmholtz instability, however, even if the discontinuities are treated numerically. All wavenumber modes are unstable to growth when the interface is discontinuous \cite{chandrasekhar61}, thus numerical noise may seed secondary instabilities that were not present in the initial conditions.  Increasing the resolution of the calculation can alter the solution obtained since this permits access to higher wavenumber modes.  Assertions have been made that some methods are of higher quality than others since they produce more detailed structure \cite{arepo, hopkins15}, however, it is inappropriate to attribute structure generated from numerical noise as corresponding to resolved substructure \cite{robertsonetal10, mlp12, lecoanetetal16}.



In this work, we study the convergence properties of SPH on the Kelvin-Helmholtz instability. We use the smooth, well-posed initial conditions of Lecoanet et al \cite{lecoanetetal16} to obtain a converged solution, avoiding the issues inherent to discontinuous initial conditions. The calculations utilise a Navier-Stokes viscosity and physical thermal conductivity, thereby permitting the calculations to convergence in resolution to a particular solution in the strongly non-linear regime. Lecoanet et al \cite{lecoanetetal16} obtained converged solutions for this problem in both the linear and strongly non-linear regime using the finite-volume grid code {\sc Athena} \cite{athena} and the pseudo-spectral code {\sc Dedalus}.

\section{Physical equations and initial conditions}

We solve the set of equations given by
\begin{gather}
\frac{{\rm d}\rho}{{\rm d}t} = - \rho \nabla \cdot {\bm v}, \label{eq:cty} \\
\frac{{\rm d}{\bm v}}{{\rm d}t} = - \frac{\nabla P}{\rho} - \frac{1}{\rho} \nabla \cdot {\bm \Pi}, \\
\frac{{\rm d}u}{{\rm d}t} = - \frac{P}{\rho}\nabla \cdot {\bm v} + \nabla \cdot ({\bm v} \cdot {\bm \Pi}) + \frac{1}{\rho} \nabla \cdot (\chi \rho \nabla T), \\
\frac{{\rm d}c}{{\rm d}t} = \frac{1}{\rho} \nabla \cdot (\nu_{\rm c} \rho \nabla c) , \label{eq:colour}
\end{gather}
where $\rho$ is the density, ${\bm v}$ is the velocity, $P$ is the pressure, ${\bm \Pi}$ is the Navier-Stokes stress tensor, $u$ is the internal energy, $T$ is the temperature, and $\chi$ is the thermal diffusivity. A passive scalar, $c$, which we call `colour', is used to quantify the mixing of the two fluid regimes. An ideal equation of state is used, $P = \rho T$, with ratio of specific heats  $\gamma = 5/3$. The temperature is related to the internal energy according to $T = (\gamma -1) u$. The material derivative is ${\rm d}/{\rm d}t \equiv \partial / \partial t + {\bm v} \cdot \nabla$. The Navier-Stokes stress tensor is given by
\begin{equation}
\Pi^{ij} = \nu \rho \left( \frac{\partial v^i}{\partial x^j} + \frac{\partial v^j}{\partial x^i} - \frac{2}{3} \frac{\partial v^k}{\partial x^k} \delta^{ij} \right) , \label{eq:NStensor}
\end{equation}
with shear viscosity $\nu$. The colour is passively advected with the flow, but includes a `physical' diffusion term analogous to thermal conductivity. The evolution of the Kelvin-Helmholtz instability in the non-linear regime is strongly sensitive to the dissipation, both numerical and physical. In order to enforce one particular solution, the dissipation is made resolution independent through the physical dissipation terms.

The initial conditions are two-dimensional and given by
\begin{gather}
\rho = 1, \\
v_x = v_0 \left[ \tanh\left( \frac{y - y_1}{a}\right) - \tanh\left( \frac{y - y_2}{a} \right) -1 \right],  \\
v_y = A \sin(2 \pi x) \left[ \exp \left( - \frac{(y - y_1)^2}{\sigma^2} \right) + \exp\left( - \frac{(y - y_2)^2}{\sigma^2} \right) \right], \\
P = 10, \\
c = \frac{1}{2} \left[ \tanh \left( \frac{y - y_1}{a} \right) - \tanh\left( \frac{y - y_2}{a} \right) + 2 \right] ,
\end{gather}
where $a = 0.05$, $\sigma = 0.2$, $A = 0.01$, $v_0 = 1$, $y_1 = 0.5$ and $y_2 = 1.5$. The calculations are performed in a periodic box of size $x \in [0,L]$ and $y \in [0, 2L]$ with $L=1$. The Reynolds number used is Re=$10^5$, defined according to
\begin{equation}
\text{Re} = \frac{L \Delta v}{\nu} , 
\end{equation}
where $\Delta v = 2 v_0$. This yields $\nu = 2 \times 10^{-5}$, and for these calculations $\nu = \chi = \nu_{\rm c}$. 

\section{Numerical method}

We use SPH to solve equations~\ref{eq:cty}--\ref{eq:colour}. We use a standard formulation of SPH where the density is computed through summation over neighbouring particles, and the smoothing length and density are self-consistently obtained through iteration \cite{monaghan05, price12}. Artificial viscosity is used with the Morris \& Monaghan limiter \cite{mm97}. The calculations use the septic (M8) spline, a high-order kernel from the same family as the cubic and quintic B-splines \cite{schoenberg46}.

The Navier-Stokes viscosity is solved using a two-first derivatives implementation, similar to \cite{flebbeetal94, watkinsetal96, ss06}. The Navier-Stokes stress tensor (equation~\ref{eq:NStensor}) is computed first, then the corresponding accelerations and heating are computed using the result. This implementation exactly conserves energy and momentum. The thermal conductivity and colour diffusion are computed directly using a second derivative in the manner of Brookshaw \cite{brookshaw85} and Cleary \& Monaghan \cite{cm99}. Total energy and colour are preserved with these schemes.

Calculations are performed using $256 \times 592$, $512 \times 1184$, $1024 \times 2364$ and $2048 \times 4728$ particles arranged on triangular lattices, labelled as $n_{\rm x}=256$, 512, 1024 and 2048, respectively. The $n_{\rm x}=2048$ calculation required $\sim 70~000$ cpu-hours of computational time.

\section{Results}

The goal is to obtain convergence of SPH calculations on the Kelvin-Helmholtz instability test of Lecoanet et al \cite{lecoanetetal16}. In absence of an analytic solution, convergence is measured by comparing SPH results to the D2048 solution obtained using the pseudo-spectral code {\sc Dedalus} at a resolution of $n_{\rm x}=2048$ \cite{lecoanetetal16}. 

\begin{figure}
\centering
\includegraphics[width=0.24\linewidth]{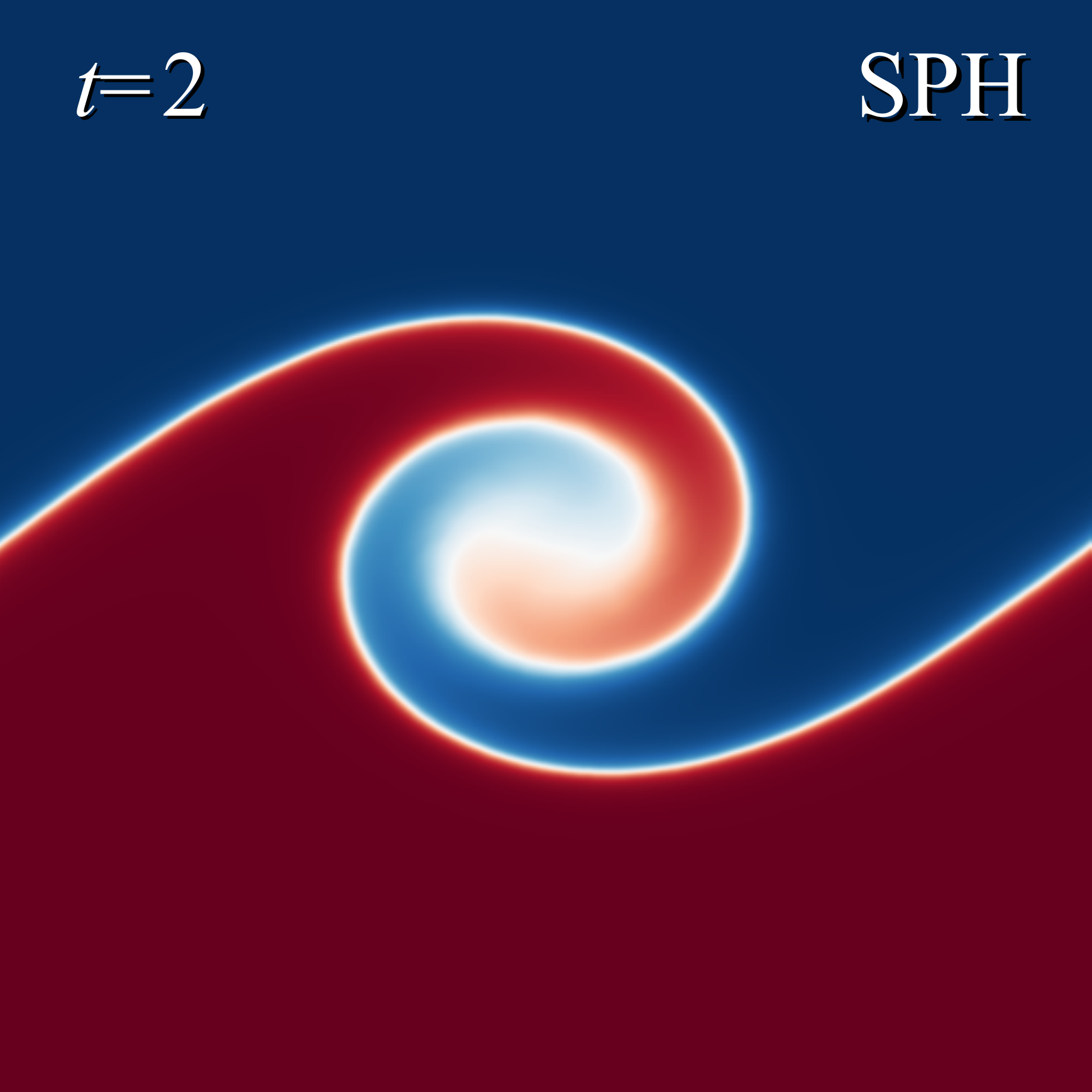} \hspace{-2.1mm}
\includegraphics[width=0.24\linewidth]{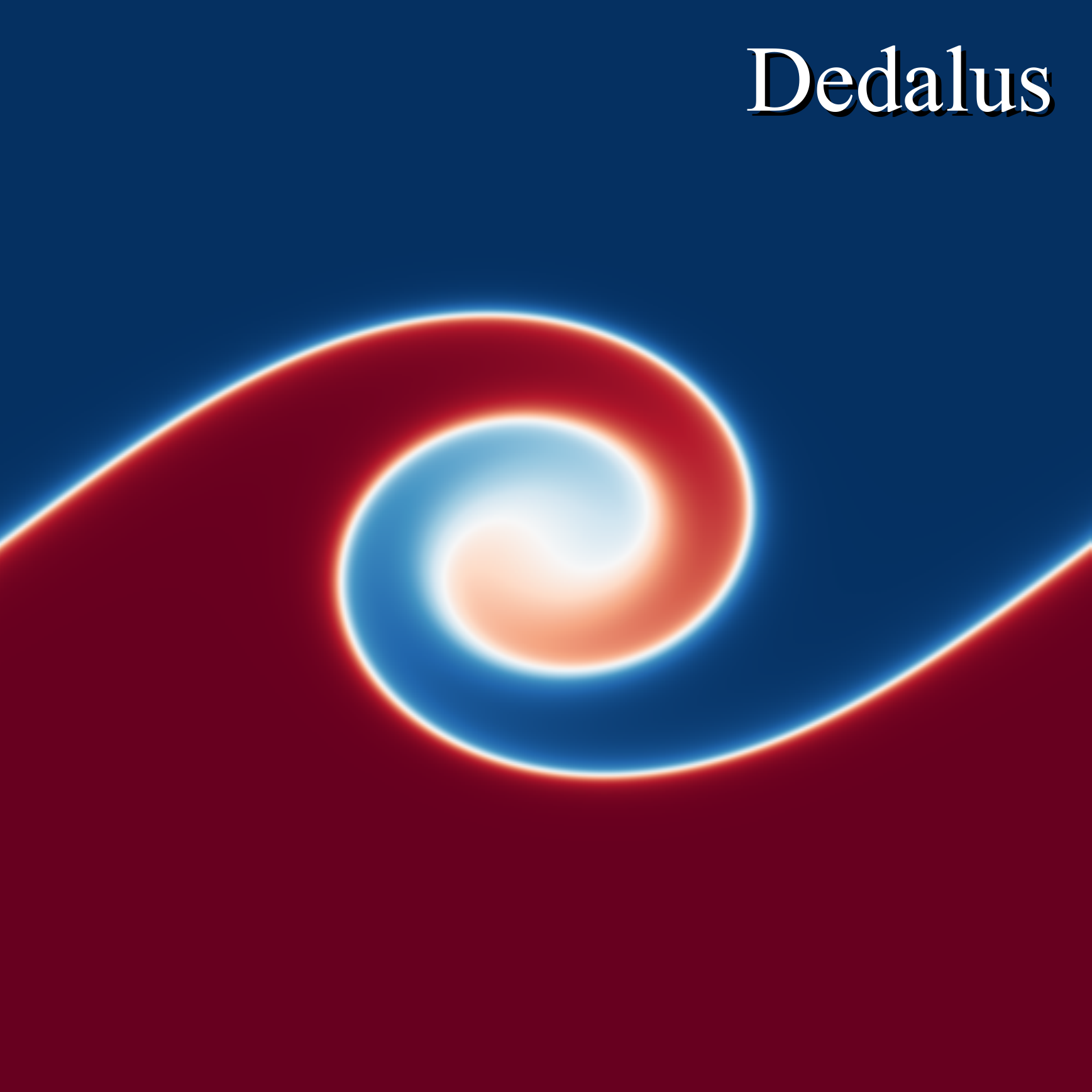} 
\hspace{2mm}
\includegraphics[width=0.24\linewidth]{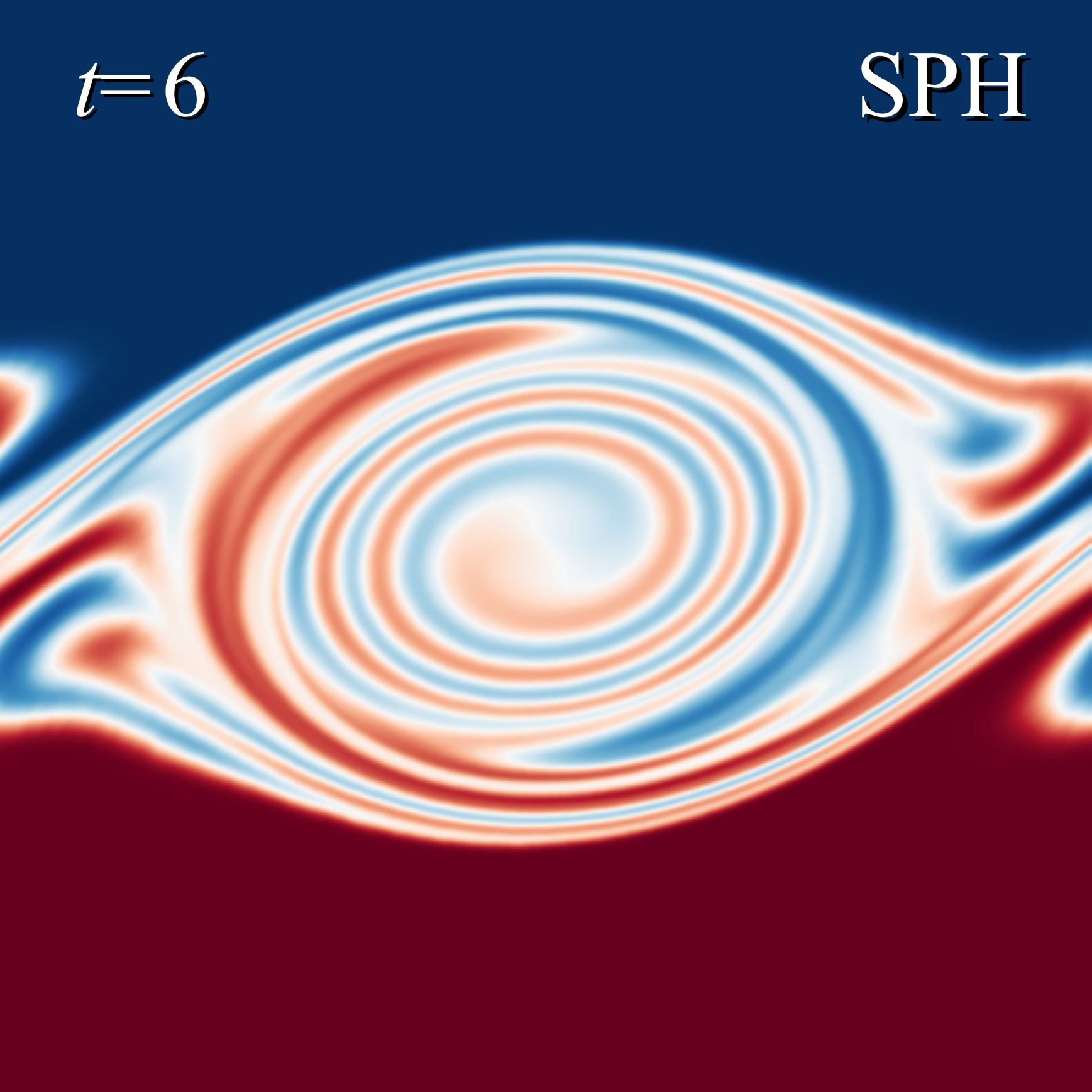} \hspace{-2.1mm}
\includegraphics[width=0.24\linewidth]{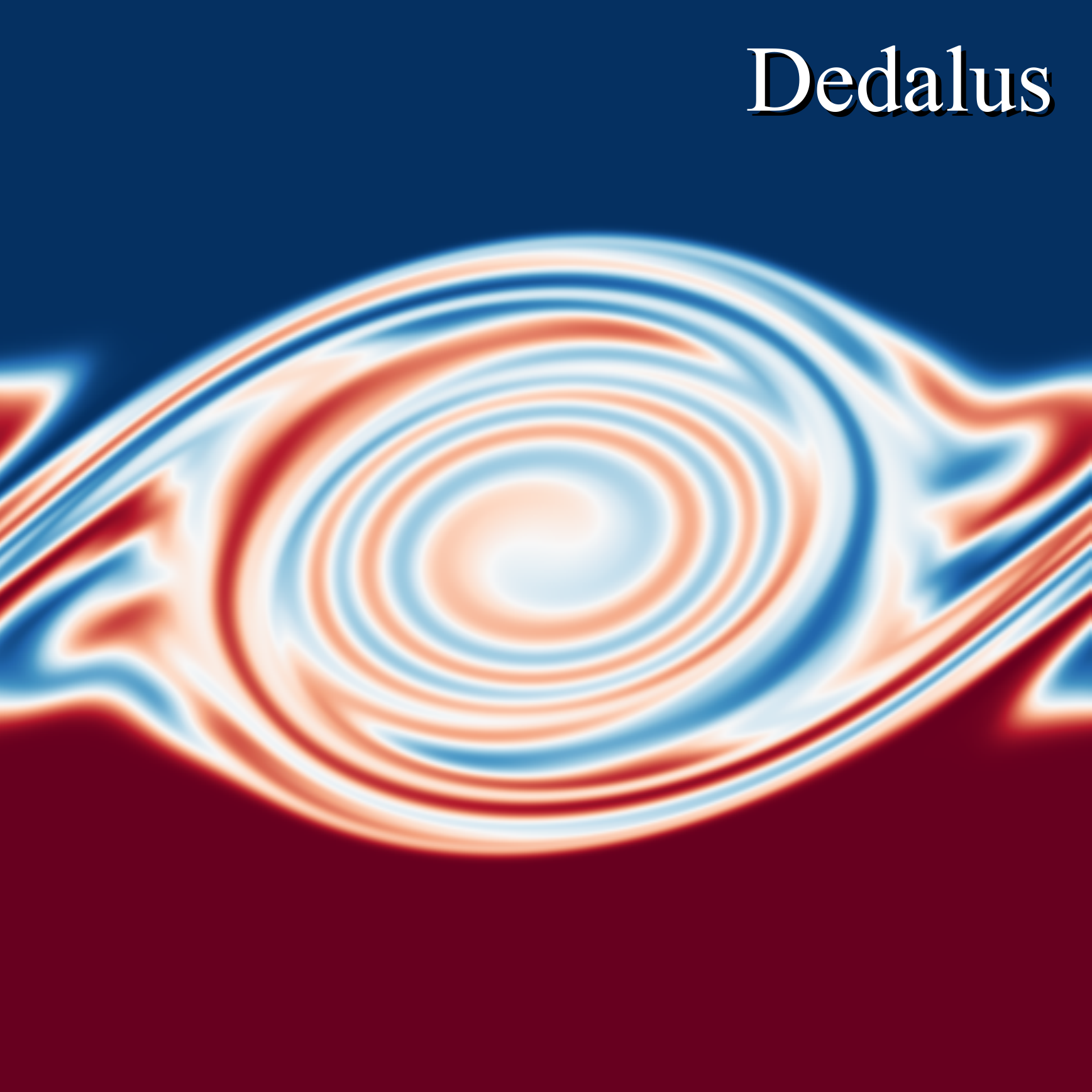} \\
\includegraphics[width=0.24\linewidth]{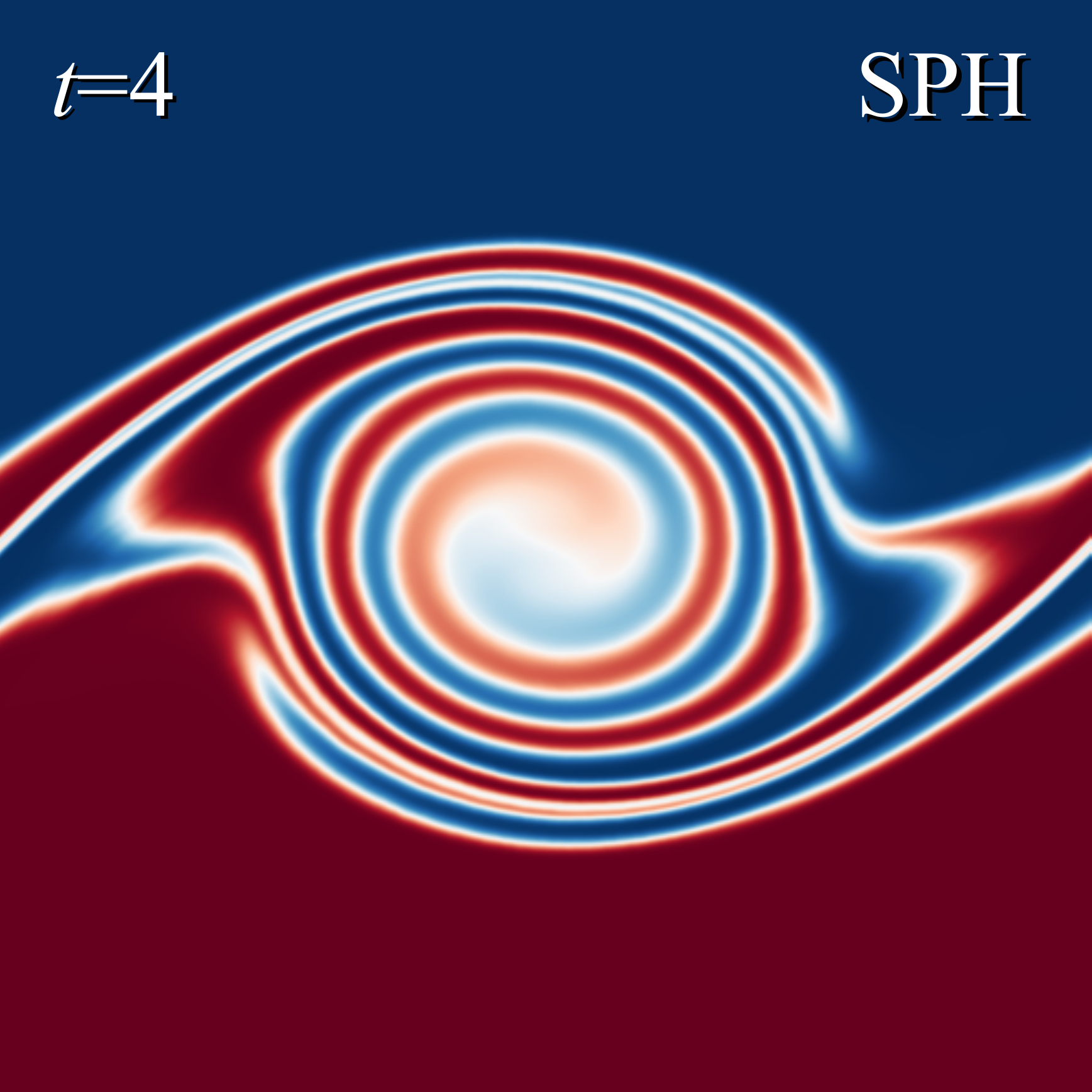} \hspace{-2.1mm}
\includegraphics[width=0.24\linewidth]{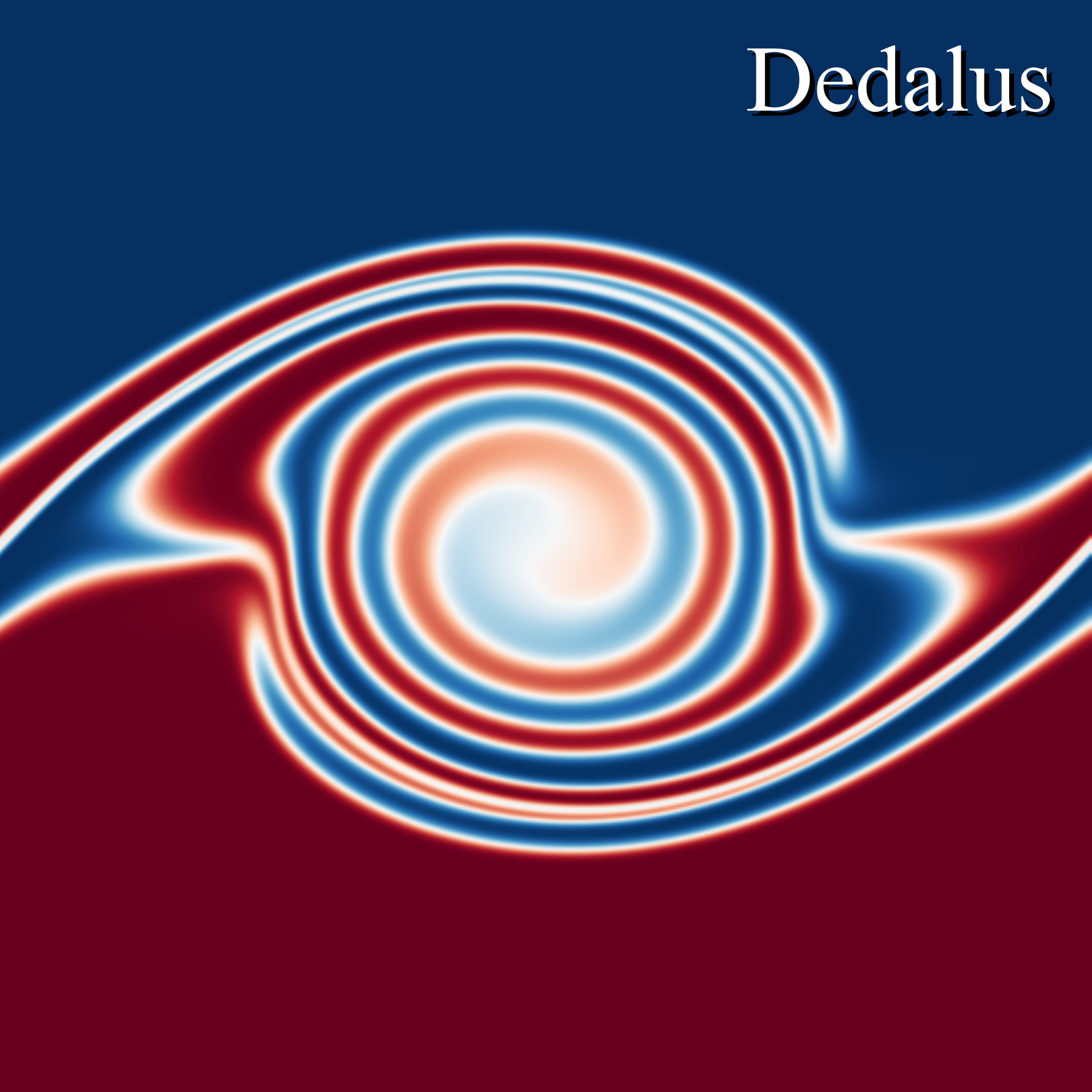} 
\hspace{2mm}
\includegraphics[width=0.24\linewidth]{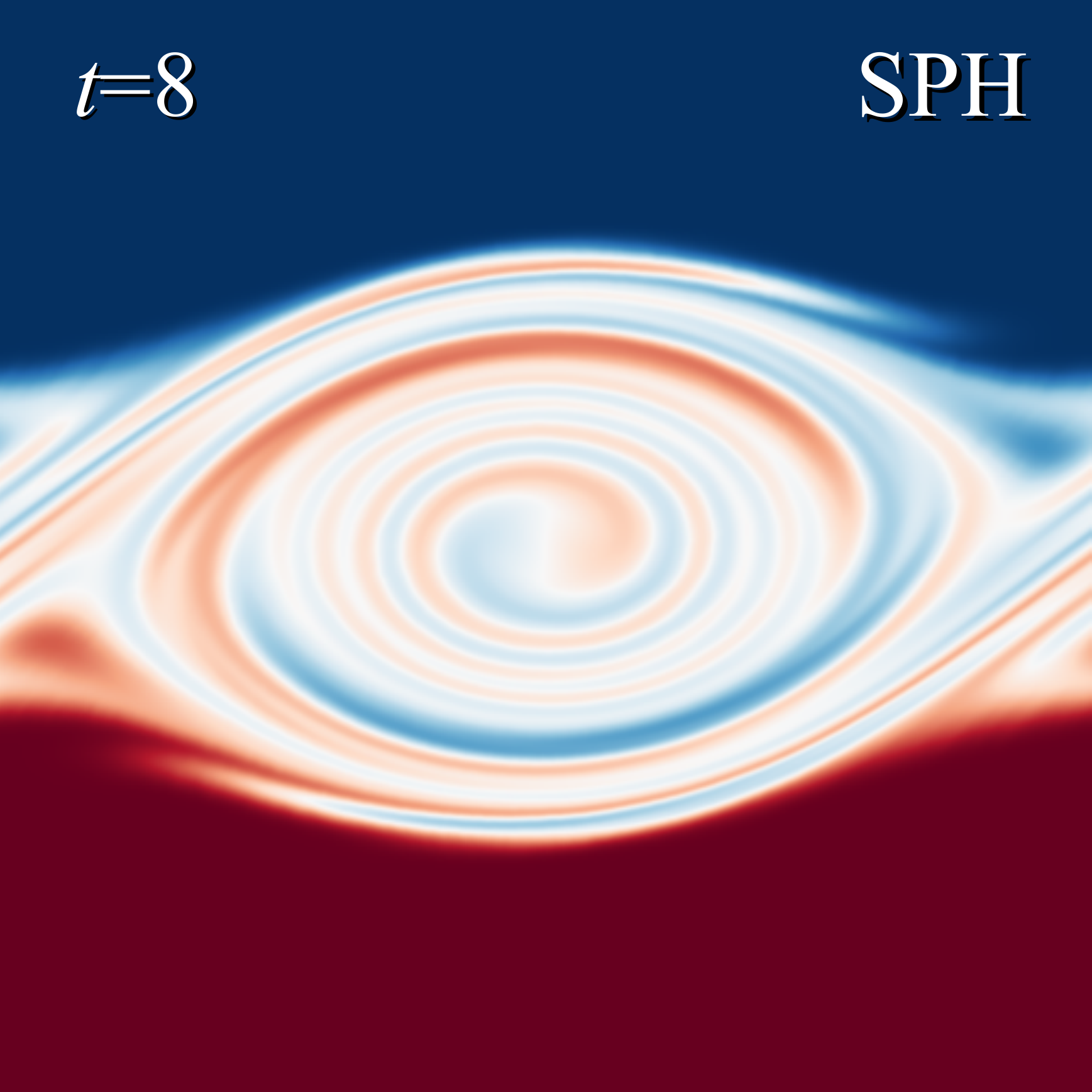} \hspace{-2.1mm}
\includegraphics[width=0.24\linewidth]{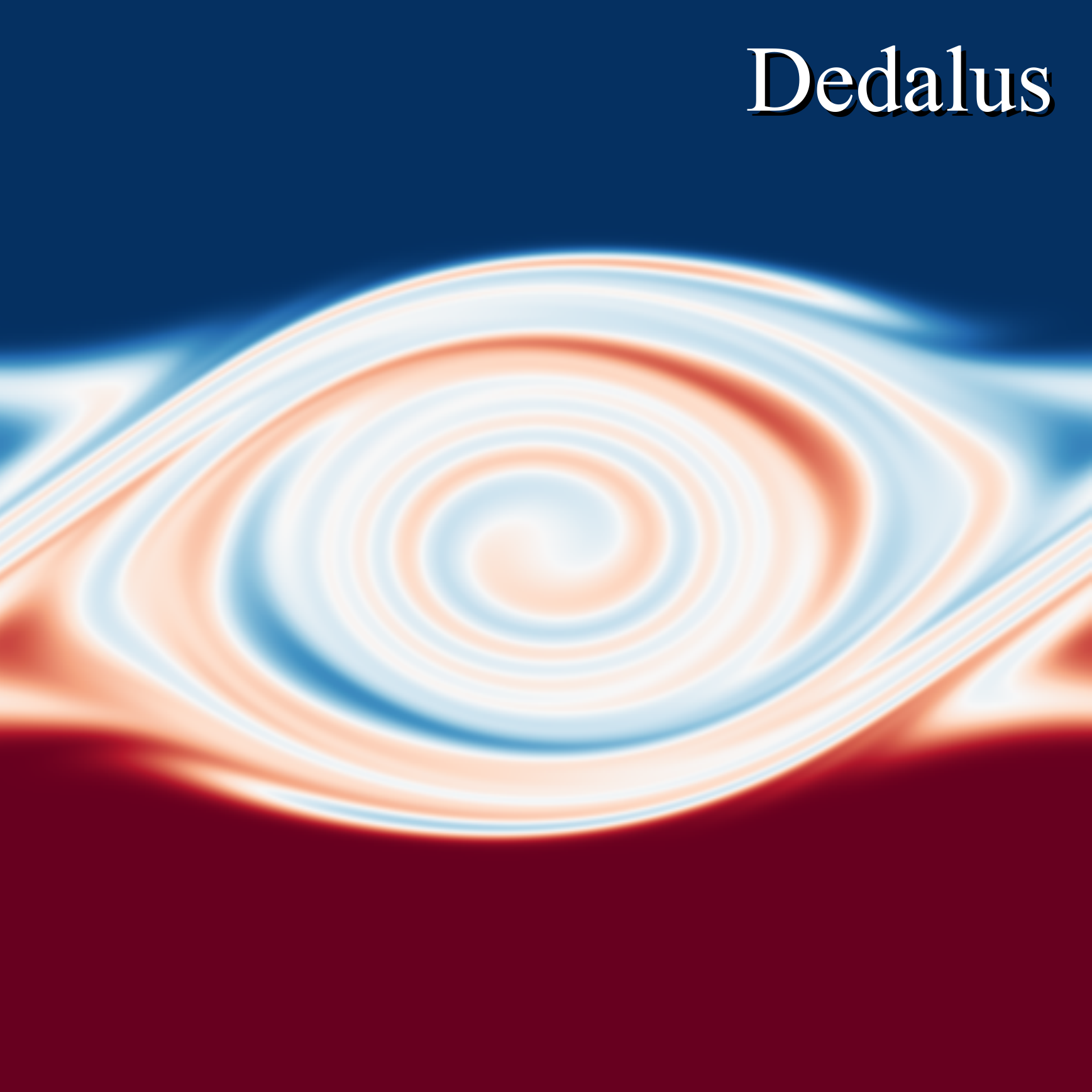} \\
\includegraphics[width=0.995\linewidth]{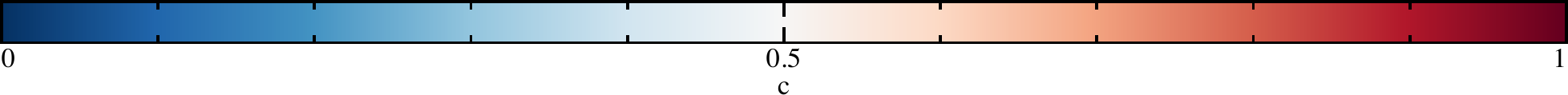}
\caption{The colour field at $t=2$, 4, 6 and 8 for the $n_{\rm x}=2048$ SPH results with the reference D2048 solution as computed using the {\sc Dedalus} code. The SPH results reproduce the reference solution at all times, though with some minor differences in the substructure in the strongly non-linear regime ($t\ge6$).}
\label{fig:render}
\end{figure}

Figure~\ref{fig:render} shows the colour field at $t=2$, 4, 6 and 8 for the $n_{\rm x}=2048$ SPH calculations alongside the D2048 reference solution. The structures in both solutions are remarkably similar at all times. In each case, the seeded mode forms a large singular curl ($t=2$), which continues to wind ($t=4$), leading to the two fluids becoming well mixed along the interface at late times ($t=6$ and 8). One minor difference is the degree of winding of the inner tip of the curl, which is not as tight for the SPH calculations compared to the reference solution. Several of the spurs at in the $t \ge 6$ snapshots are also of slightly different lengths. Despite these small differences, the SPH solution is in close proximity to the reference solution.

\begin{figure}
\centering
\includegraphics[width=0.55\linewidth]{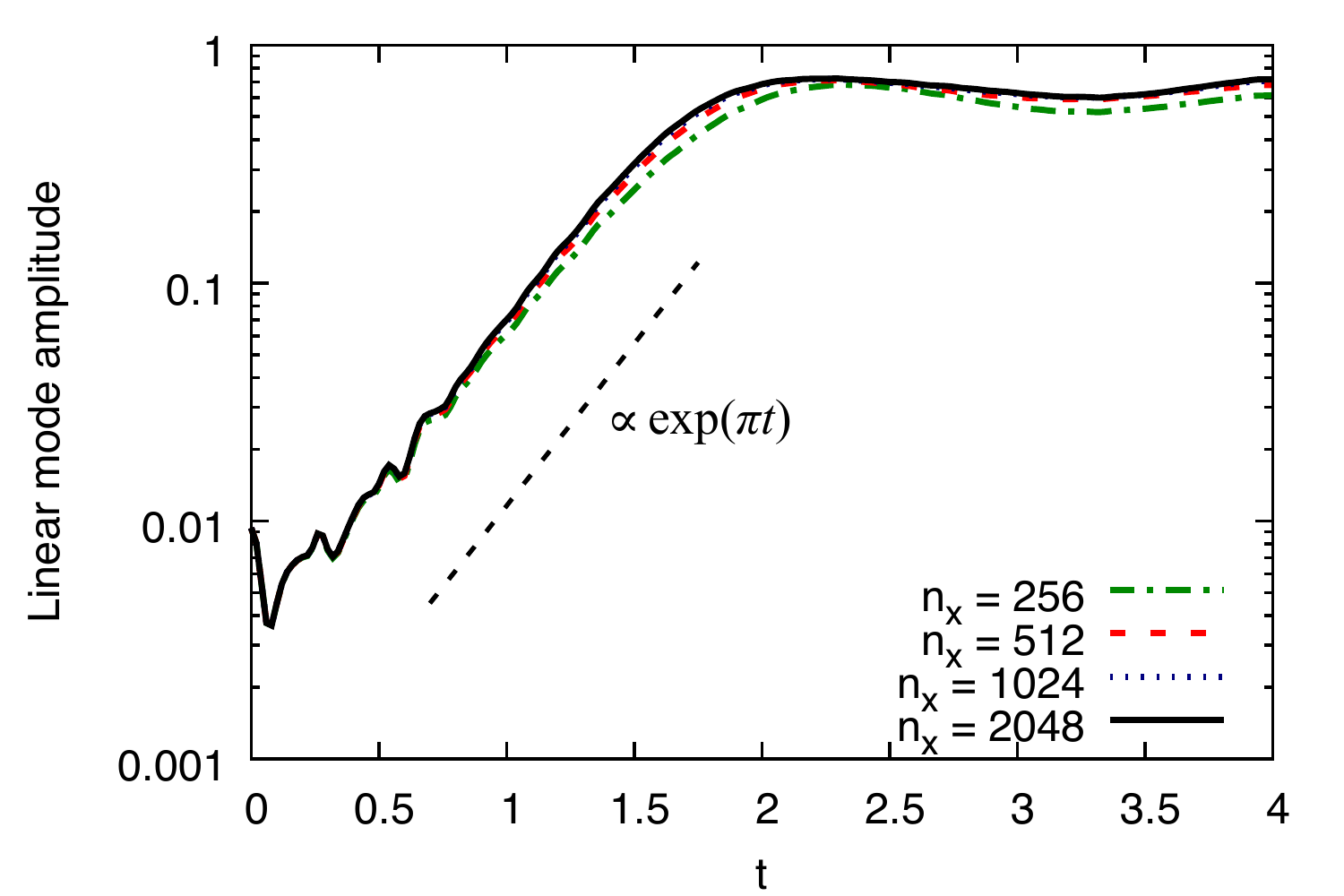}
\caption{Growth of the seeded mode of the Kelvin-Helmholtz instability. The mode amplitude growth rate is converged to $\propto \exp(\pi t)$ for all resolutions tested.}
\label{fig:modeamp}
\end{figure}

Figure~\ref{fig:modeamp} shows the growth of the amplitude of the seeded mode in the linear regime, obtained in a manner according to McNally, Lyra \& Passy \cite{mlp12}. The measured growth rate of the mode is converged to $\propto \exp(\pi t)$. It is difficult, however, to obtain an analytic estimate of the growth rate for this problem. An incompressible fluid with discontinuous interfaces should have a growth rate that is $\propto \exp(2 \pi t)$ \cite{chandrasekhar61}. A smoothed velocity interface will reduce the growth rate by approximately 20\% \cite{wyl10}. It is reasonable to expect that the growth rate for this problem should be even slower than that. These calculations are for a compressible fluid, not incompressible, and additionally smooth the initial velocity perturbation used to seed the instability. Thus, while no analytic estimate of the growth rate of this mode is available, the results are at least consistent with the estimates available acting as upper bounds. The growth rate is converged even for our lowest resolution of $n_{\rm x}=256$ particles.

\begin{figure}
\centering
\includegraphics[width=0.55\linewidth]{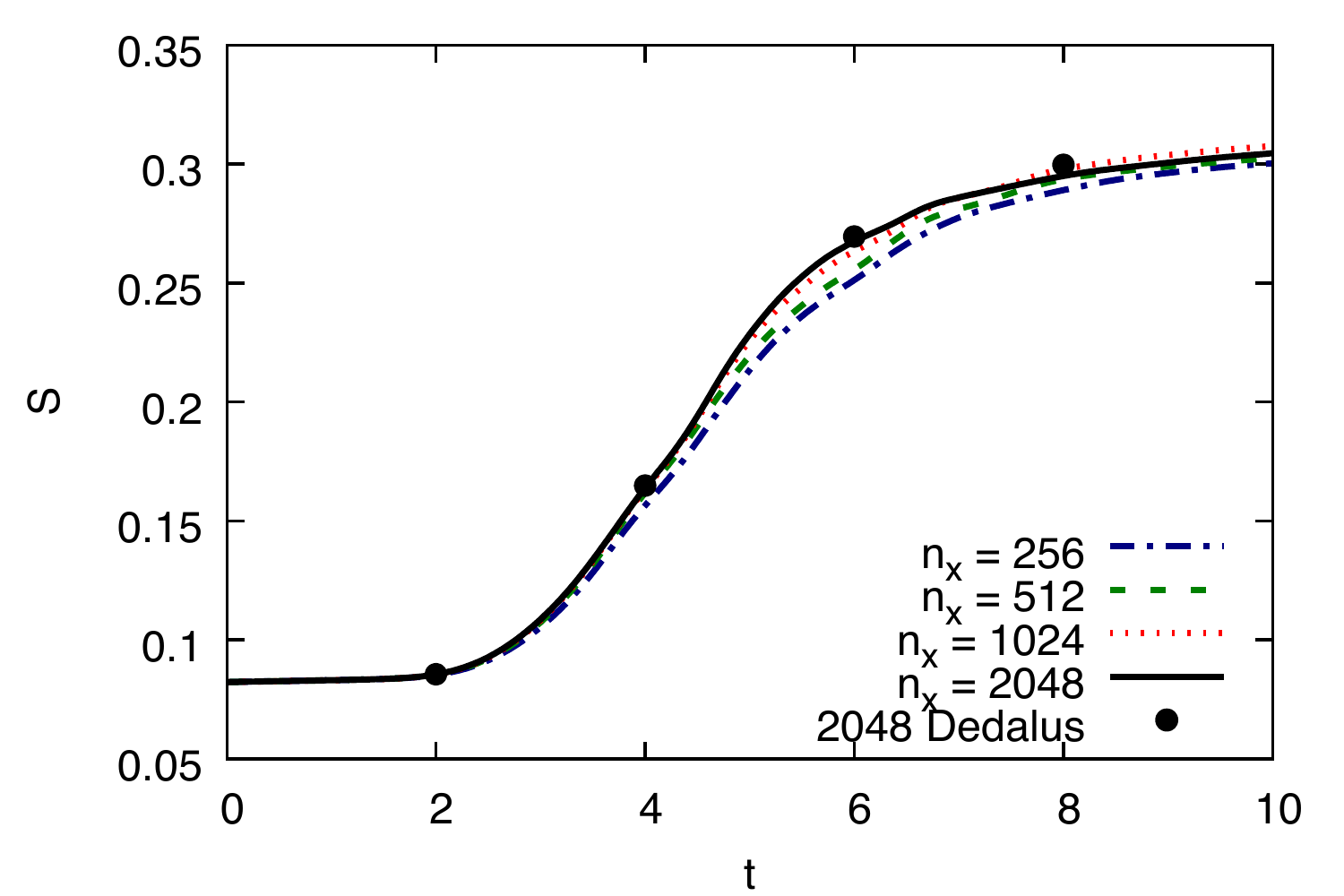}
\caption{Total colour entropy as a function of time. Black circles are values from the reference solution. The SPH calculations reproduce the magnitude and shape of the total colour entropy from the reference solution.}
\label{fig:ce}
\end{figure}

The degree is mixing is quantified by an entropic quantity. Defining
\begin{equation}
s = - c \ln(c)
\end{equation}
as the colour entropy, the total colour entropy is calculated by the volume integral
\begin{equation}
S = \int \rho s {\rm d}V ,
\end{equation}
which may computed in SPH by the summation $\sum_a m_a s_a$. The total colour entropy only increases when $\nu_{\rm c} > 0$. Figure~\ref{fig:ce} shows the total colour entropy as a function of time, with black circles the values of the reference solution at $t=2$, 4, 6 and 8. The shape of the total colour entropy curve matches the reference solution (see also \cite{lecoanetetal16}), and is converging towards the reference data points as the resolution improves.

\begin{figure}
\centering
\includegraphics[width=0.55\linewidth]{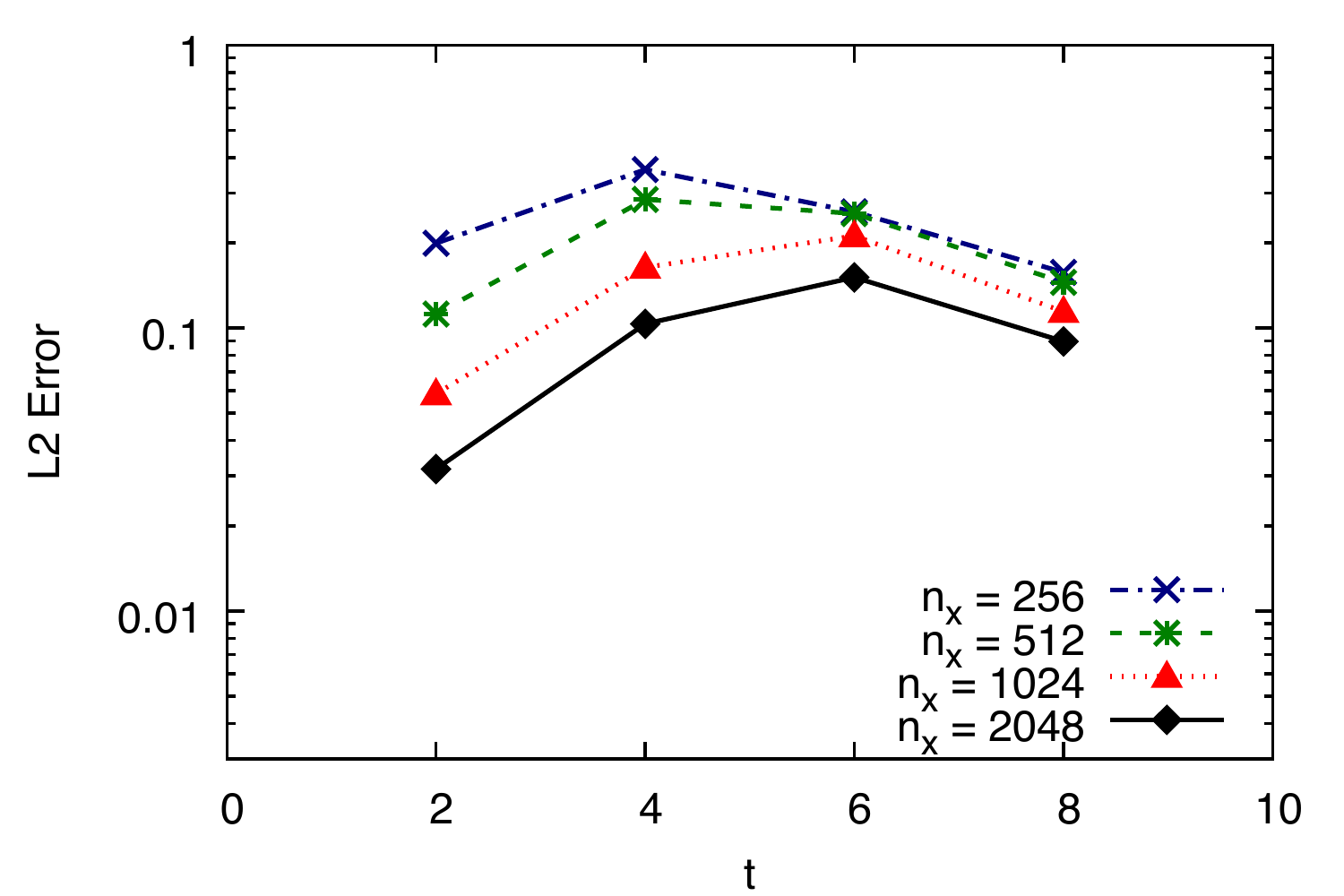}
\caption{$\mathcal{L}_2$ errors of the SPH calculations with respect to the reference solution. The convergence is linear in resolution at $t=2$. The convergence rate is slower in the strongly non-linear regime, though the errors continue to reduce, converging to the reference solution.}
\label{fig:l2err}
\end{figure}

Formal convergence of our calculations is assessed through the $\mathcal{L}_2$ error. The SPH particles for each calculation are interpolated to a $2048 \times 4096$ grid and compared grid-cell to grid-cell with the reference solution. The error is computed according to
\begin{equation}
\mathcal{L}_2 = \left[ \sum_a (c_a^{\rm SPH} - c_a^{\rm D2048})^2 {\rm d}V \right]^{1/2} ,
\end{equation}
where ${\rm d}V = 2048^{-2}$ is the volume of each grid-cell, and $c^{\rm SPH}$ and $c^{\rm D2048}$ are the colour fields of the gridded SPH data and reference solution, respectively. 

\begin{table}
\caption{\label{tbl:convergence}The measured convergence rates, $\Gamma$, of the $\mathcal{L}_2$ error as fit to $\propto n_{\rm x}^{- \Gamma}$.}
\centering
\begin{tabular}{cc}
\br
$t$  & $\Gamma$ \\
\mr
2 & 0.89 \\
4 & 0.62 \\
6 & 0.26 \\
8 & 0.28 \\
\br
\end{tabular}
\end{table}

Figure~\ref{fig:l2err} shows the $\mathcal{L}_2$ error of the SPH calculations with respect to the reference solution at $t=2$, 4, 6 and 8. Table~\ref{tbl:convergence} lists the measured convergence rates fit to $\propto n_{\rm x}^{-\Gamma}$. The errors in the SPH calculations may be contrasted to Athena calculations in Figure~2 of \cite{lecoanetetal16}. The maximum errors of the SPH calculations are of order $10^{-1}$, whereas the A1024 and A2048 calculations have errors on the order of $\sim10^{-3}$ and $\sim10^{-4}$, respectively, for $t > 4$.

The error reduces linearly with respect to resolution at $t=2$, in line with the expected rate of convergence from the artificial dissipation. Even for the highest resolution calculation, the dissipation of kinetic energy from the artificial dissipation remains comparable to the dissipation by the Navier-Stokes viscosity. The convergence rate becomes sub-linear in the non-linear regime ($t \ge 4$), caused primarily by errors in the pressure gradient, which scale as $\mathcal{O}(1)$ with respect to resolution. This may be contrasted to {\sc Athena} calculations, which have a measured rate of convergence that is between second to third order \cite{lecoanetetal16}. Importantly, the error of the SPH calculations is reducing at all times once the resolution is greater than $n_{\rm x}=512$ particles. 

Reducing the error in the pressure gradient can be done by using higher-order smoothing kernels. For this problem, it was found that to obtain high-quality results necessitated the use of the septic spline, a high-order smoothing kernel from the B-spline family of kernels \cite{schoenberg46}. The septic spline is the high-order relative of the commonly used cubic and quintic splines. Using a high-order kernel is not an intrinsic requirement to activate the Kelvin-Helmholtz instability, but rather is needed to capture the amplitude of the initial velocity perturbation in the initial conditions and reduce spurious growth of other modes. 

\begin{figure}
\centering
\includegraphics[width=0.196\linewidth]{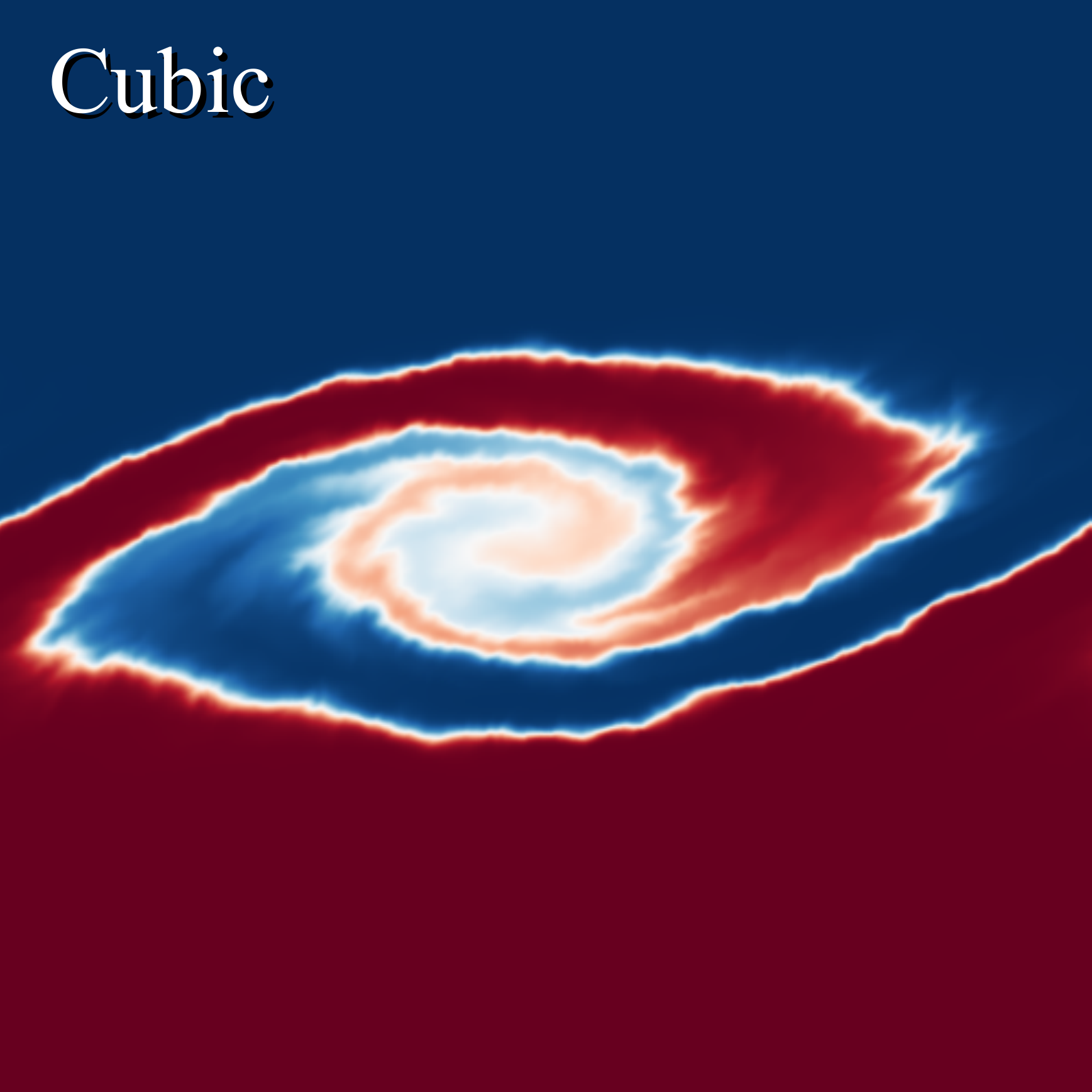} \hspace{-2mm}
\includegraphics[width=0.196\linewidth]{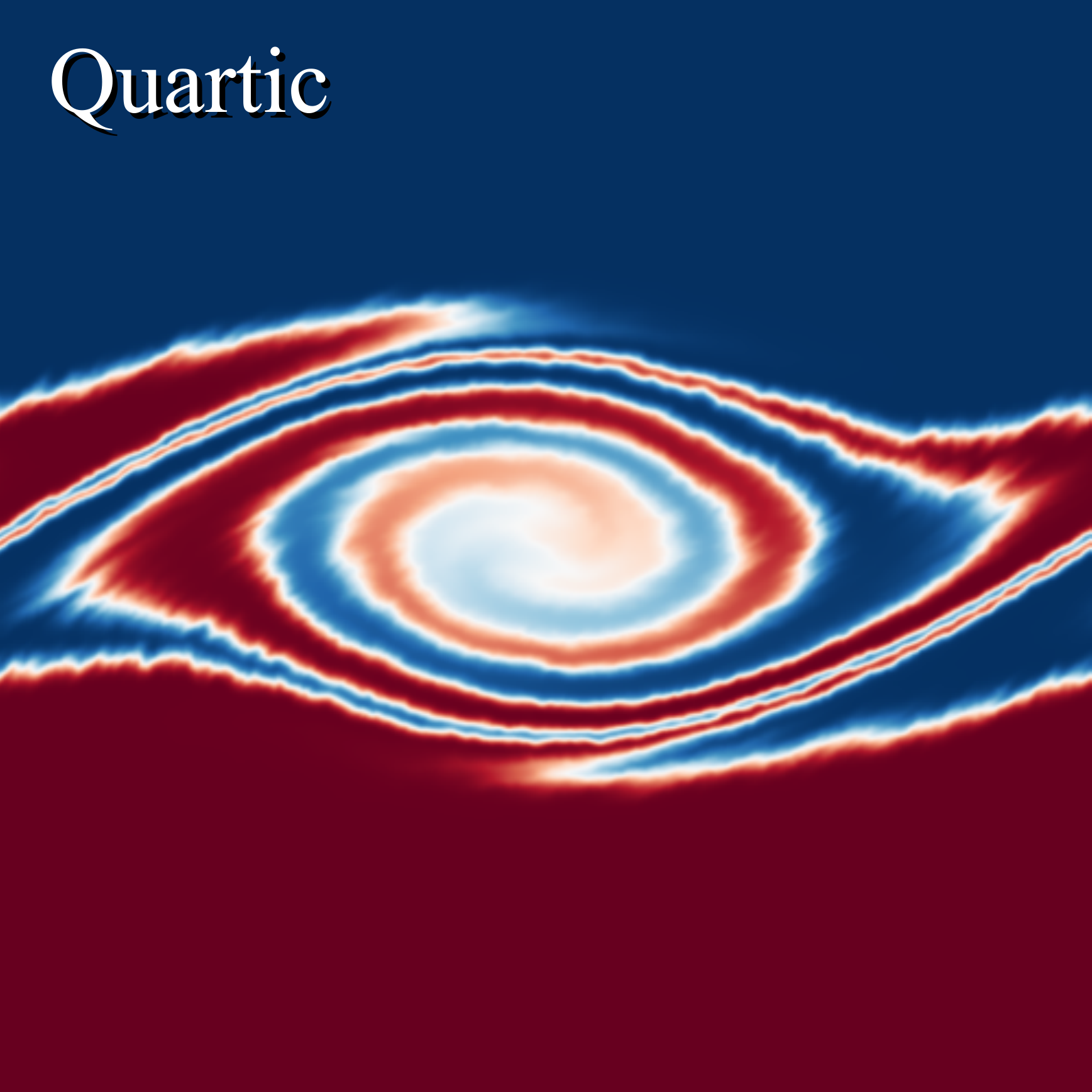} \hspace{-2mm}
\includegraphics[width=0.196\linewidth]{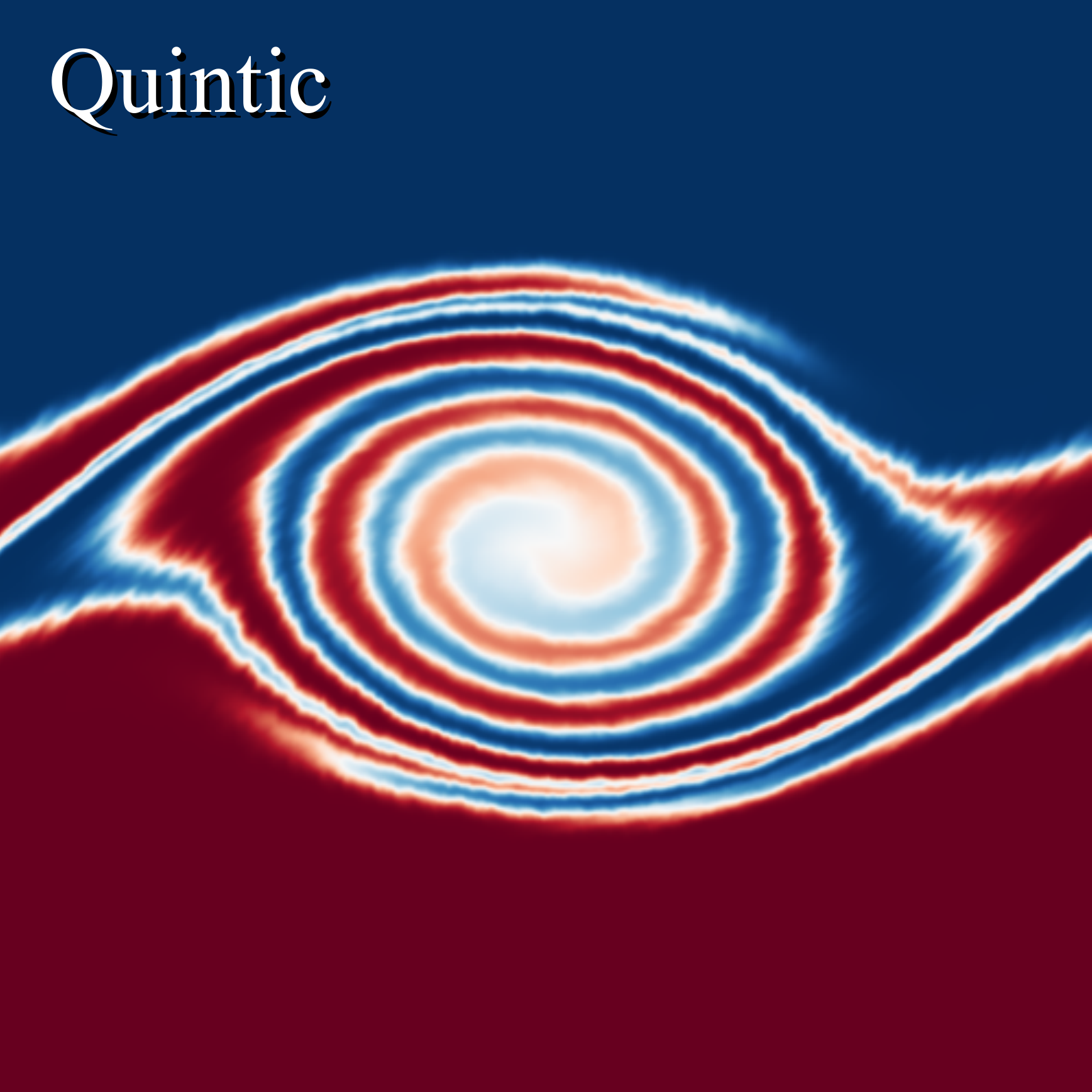} \hspace{-2mm}
\includegraphics[width=0.196\linewidth]{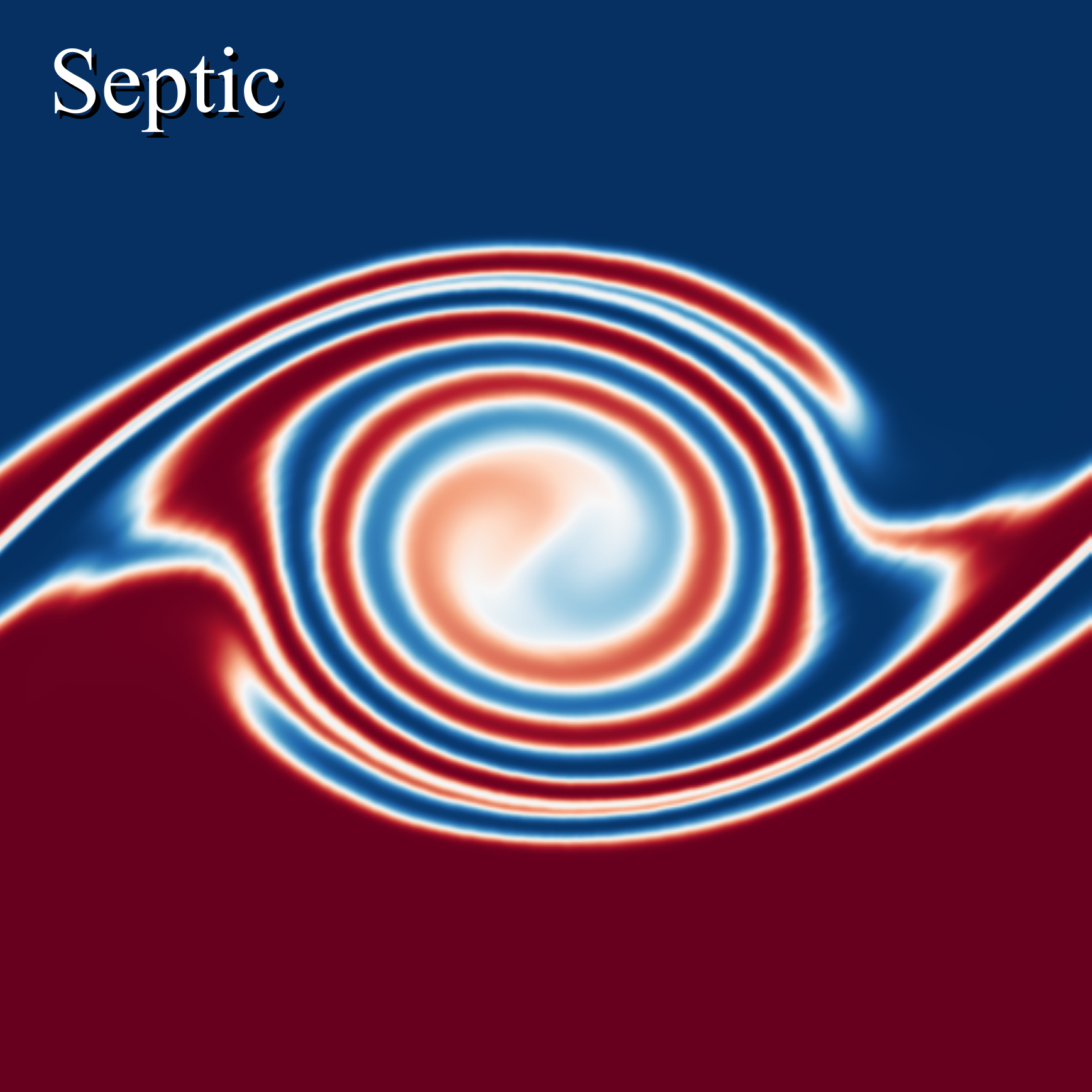} \hspace{-2mm}
\includegraphics[width=0.196\linewidth]{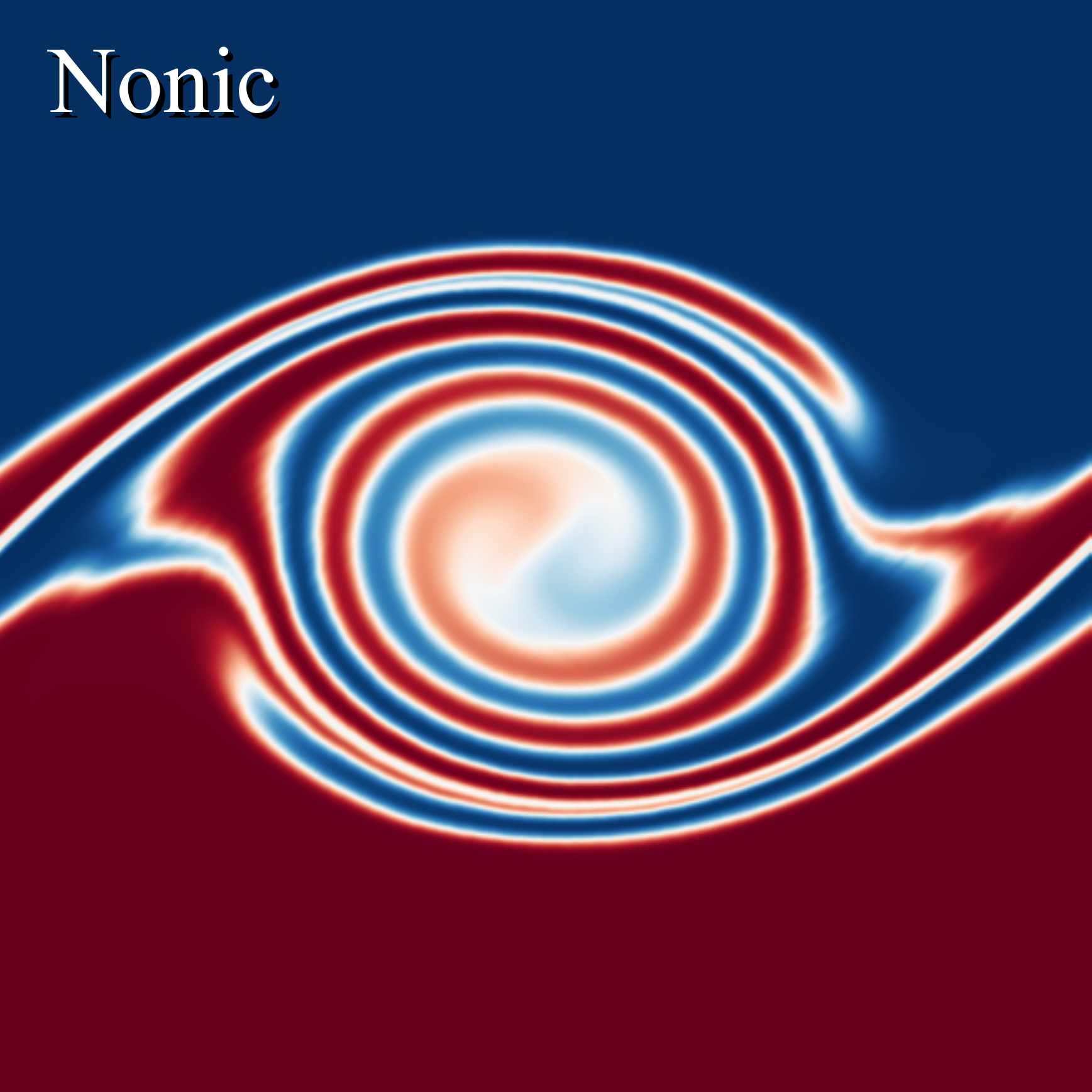} \\
\includegraphics[width=0.995\linewidth]{cbar-wide.pdf}
\caption{The colour field at $t=4$ for SPH calculations using the cubic to nonic splines. For low-order kernels, they become the dominant source of error which degrades the quality of solution obtained. For high-order kernels (septic, nonic), the quality of smoothing kernel is no longer the dominant source of error and the solutions are indistinguishable. The calculations presented use the septic spline, which is the minimum order of B-spline that does not influence the results.}
\label{fig:splines}
\end{figure}

Figure~\ref{fig:splines} show the colour field at $t=4$ for calculations using the cubic spline through to nonic spline. For the cubic and quartic spline, the evolution of the vortex is significantly altered with respect to the reference solution (c.f.~Figure~\ref{fig:render}). The resemblance of the SPH colour fields to the reference solution improves as the quality of smoothing kernel improves. The difference between the septic and nonic spline results is negligible, thus we conclude that the kernel bias is a sub-dominant source of error when these high-order kernels are used. The calculations examined in this work have used the septic spline.


\section{Summary}

We have performed SPH calculations of the Kelvin-Helmholtz instability. We used the smooth, well-defined initial conditions of Lecoanet et al \cite{lecoanetetal16}, thereby avoiding convergence issues with discontinuous initial conditions. A Navier-Stokes viscosity and thermal conductivity were included to force the evolution of the instability in the non-linear regime. A passive scalar, referred to as `colour', was added to the two fluids to measure the degree of mixing. A high-resolution calculation using the pseudo-spectral code {\sc Dedalus} provided a reference solution to which results were compared. 

The SPH results qualitatively matched the reference solution. The two fluids formed a curled vortex along the interface, which continued to spiral producing substructure and mixing of the two fluids. The substructure generated within the vortex matched between the SPH calculations and the reference solution well into the late non-linear regime.

Quantitative numerical convergence was measured for SPH on the Kelvin-Helmholtz instability, albeit with only linear convergence in the $\mathcal{L}_2$ error in the linear regime of the instability ($t=2$), and sub-linear convergence in the strongly non-linear regime. The degree of mixing was measured by defining an entropic quantity for the `colour' field, and the total entropy of the colour field increased monotonically in agreement with the curve from the reference solution.

Importantly, convergence has been demonstrated using the standard form of SPH with an artificial viscosity, of the kind that has been used for decades \cite{monaghan92, monaghan05}. No alternative SPH formulations, modifications, or hacks were employed. The only requirement to achieve the results presented here was a high-order smoothing kernel, which was needed in order to capture the initial amplitude of the velocity perturbation in the initial conditions and to curb growth of other modes from numerical noise. We conclude that the rumours of the `fundamental flaws' of SPH have been grossly exaggerated. 

\ack

This work was motivated by attendance at AstroNum 2017 in Saint-Malo, France. Thank you to Daniel Lecoanet for providing the reference {\sc Dedalus} solution, and to Daniel Price for useful discussions on physical dissipation terms in SPH. 

\section*{References}

\bibliographystyle{iopart-num}
\bibliography{bib}

\end{document}